\documentclass[aps,prl,reprint]{revtex4-1}
\usepackage{amsmath}
\usepackage{amssymb}
\usepackage{graphicx}
\usepackage{mathrsfs}
\usepackage{subfigure}
\usepackage{upgreek}

\newcommand{ \braces} [1] { \left( #1 \right)}
\newcommand{ \sbraces} [1] { \left\lbrace #1 \right\rbrace}
\newcommand{ \ebraces} [1] { \left[ #1 \right]} 
\newcommand{ \sech} [1] {\, \mathrm{sech} \! \left( #1 \right)}
\newcommand{ \csch} [1] {\, \mathrm{csch} \! \left( #1 \right)}
\newcommand{\sket}[1]{| #1 \rangle}
\newcommand{\ket}[1]{\left|#1\right\rangle}
\newcommand{\von} [1] {\! \braces{#1}}
\newcommand{\matr} [4] {\begin{pmatrix} #1 & #2 \\ #3 & #4 \end{pmatrix}}
\newcommand{\svek} [2] {\begin{pmatrix} #1 \\ #2 \end{pmatrix}}
\newcommand{ \infinteg} [1] { \int\limits_{-\infty}^\infty \! \mathrm{d}#1 \,}
\newcommand{ \infintx} { \int\limits_{-\infty}^\infty \! \mathrm{d}X \,}
\newcommand{\abs} [1] {\left| #1 \right|}

\begin{document}

\title{An atomic bright vector soliton as an active particle}

\author{Timo Eichmann}
\author{James R. Anglin}

\affiliation{\mbox{State Research Center OPTIMAS and Fachbereich Physik,} \mbox{Technische Univerit\"at Kaiserslautern,} \mbox{D-67663 Kaiserslautern, Germany}}

\date{\today}

\begin{abstract}
Solitons in general are configurations of extended fields which move like isolated particles. Vector bright solitons can occur in a two-component self-attractive Bose-Einstein condensate. If the components of the condensate have different chemical potentials, the total spin of the soliton can serve as an internal energy depot that makes the soliton into an \emph{active} particle, able to move against an external force using energy carried within the particle---if there is a dynamical mechanism for steadily transferring energy from soliton spin into soliton motion. Here we present such a dynamical mechanism, embed it in an experimentally feasible way within the larger system of a spinor condensate mean field, and show how the mechanism works to realize a solitonic active particle. In what can be considered a toy model for the project of going beyond toy models for active particles, we test the robustness of the activity mechanism by exploring a range of deformations to the simplest model for embedding the nonlinear mechanism in the condensate system.
\end{abstract}

\maketitle

\section{I. Introduction}

One of the most elementary tasks in nature and everyday life is the transport of objects. Nature has developed `motor proteins' to achieve transport in cells \cite{Kinesin1, Kinesin2}, while humans build engines for larger scale transport. In both cases some kind of internal energy is used to perform motional work (including work against friction). The phenomenological model of an \emph{active particle} \cite{Active1, Active2, Daemon4, Active3, Active4, Active5} has emerged as a useful concept for modelling internally powered motion in general. Fuel does not just automatically convert itself into work, however. Macroscopic engines are non-trivial dynamical mechanisms, subject to laws of thermodynamics whose emergence from microscopic mechanics is still being studied, and the microscopic mechanisms of biological motors are not yet understood. In this paper we contribute a concrete example of an active particle realized within an extended environment yet described microscopically. 

A bright matter-wave soliton is a small cloud of cohesive gas \cite{Khaykovich, Strecker, BStheo1, BStheo2, BStheo3} that has mechanical properties like a particle. External force can accelerate the small cloud as a whole, and its motion is well described with a single collective coordinate that can carry both kinetic and potential energy. If the soliton is created in a multi-component Bose-Einstein condensate \cite{SpinExp0, SpinExp1, SpinExp2, SpinExp3, SpinTheo1}, it can also have internal energy. If for example the bright vector soliton is placed in a constant magnetic field, and is composed of atoms of one atomic species with atoms in spin states $\sket{-}=\sket{F^{(1)},m_F^{(1)}}$ or $\sket{+}=\sket{F^{(2)},m_F^{(2)}}$, then the soliton has a large range of internal energies depending on the total spin of all its constituent atoms. By the experimentally feasible technique of Rabi driving, this internal energy can be coupled to the soliton's motion in such a way as to realize an active particle, within a real, extended system that can nevertheless be understood in full dynamical detail.

\subsection{I.1 The coupling problem}
Taking data from \cite{Khaykovich}, we can assume a soliton of about 6000 atoms of $^7$Li, having a total mass of $\sim 7 \cdot 10^{-26} \text{ kg}$ and a length of $\sim 2 \text{ }\mathrm{\upmu m}$. It requires roughly $1.4 \cdot 10^{-30}\text{ J}$ to lift this soliton vertically by its own length. If as in \cite{Khaykovich} we allow the soliton to be in a magnetic field of $\sim 500 \text{ G}$, a single atom that changes its state from $\sket{1,1}$  to $\sket{1,0}$ would change the internal energy of the soltion by $\sim 5 \cdot 10^{-26} \text{ J}$, which is enough energy to lift the soliton $\sim 3000$ times its own length against gravity. With no less than 6000 atoms in the soliton, it is clear that the soliton's total spin can easily store vastly abundant energy for active soliton motion.

It is also clearly possible to couple a vector soliton's spin to its collective motion. Examples using a spin-orbit coupling \cite{SolMov1} or a Rabi coupling \cite{SolMov2} show how this can be done. The `active particle' model presents the challenge, however, of non-destructively moving a soliton over a potentially long distance, against an external force, using an on-board energy depot. Oscillations of the soliton as in \cite{SolMov1}, which for large Zeeman frequency will have high frequency and small amplitude, do not meet this challenge; neither does the limited average motion of a spreading soliton shown in \cite{SolMov2}. Indeed the dynamical problem of coupling our vast internal spin energy into steady motional work is not merely one of selecting an arbitrary mechanism out of infinitely many possibilities. On the contrary, it is not immediately clear that any such mechanisms can even exist. 

The very abundance of the internal energy is the root of the problem. The internal spin energy is large because the Zeeman frequency $10^{-26}$J$/\hbar = 10^{9} \text{ s}^{-1}$ is about a million times higher than the kHz frequencies that are typical of motional collective modes in trapped condensates. There is no trivial way to achieve the extreme downconversion of splitting one quantum of energy at GHz frequency into a million kHz quanta which can be fed efficiently into collective motion of the soliton, steadily over a long time scale. Instead most couplings between soliton spin and soliton motion will merely be adiabatically suppressed \cite{BOA1, BOA2} in the regime of high Zeeman frequency, and do essentially nothing. The fact that high-frequency degrees of freedom may dress and renormalize low-frequency sectors, but not steadily pump energy into them, is after all a basic principle throughout modern physics \cite{RG1, RG2, RG3}.

There do exist loopholes in this principle, however, and one is to trade off length and time scales in a Chirikov resonance \cite{Chirikov,Daemon1}. In this paper we show how to implement such a `Hamiltonian daemon' Chirikov resonance engine using a Rabi coupling between the internal states $\sket{+}$ and $\sket{-}$ of atoms in a vector bright soliton, by letting the Rabi coupling depend periodically on the space coordinate along which the soliton is to be lifted. This spatially periodic Rabi coupling will effectively act as a ladder, up which the vector soliton can climb under its own active power.

\subsection{I.2 Paper outline}
In section II we will define this spatially periodic Rabi coupling \cite{Rabi} and show how it provides a Chirikov resonance \cite{Chirikov} to enable energy flow from the internal to the mechanical energy of the soliton. We will then use the Gross-Pitaevskii equation \cite{GPE} to describe the evolution of the soliton classically, showing numerical results for the motion of the soliton and the time evolutions of its internal and motional energies. We use the Split-Step Fourier method with an adaptive error control for our numerical simulations \cite{Num1, Num2}.

After thus confirming that the solitonic active particle can indeed work, we pause in section III to relate the one-dimensional Gross-Pitaevskii classical field theory to the two-degree-of-freedom Hamiltonian daemon system of \cite{Daemon1}. This approximate mapping can be used to estimate parameter ranges for different possible behaviors of the active soliton, and to explain some non-trivial features of its (in general) more complex dynamics. Our results in this section use the collective coordinate variational approach \cite{Var1}, which is a standard technique for the analytical treatment of solitons \cite{Var2, Var3, Var4, Var5, Var6, Var7, RabiSwitch, ResonantControl}.

Having seen how the ideal active soliton model succeeds in incorporating the Chirikov engine in the condensate mean field, in section IV we explore embedding the daemon mechanism in the field theory in more complex ways. In particular we consider couplings which introduce multiple Chirikov resonances, or broader ones. Since additional Chirikov resonances turn out to allow soliton motions at different speeds, we also examine the possible case of `negative' running speed, in which instead of being lifted against gravity the soliton runs down a potential slope at constant speed, using its internal spin as a brake. Finally we analyze a range of cases in which the Gross-Pitaevskii nonlinear Schr\"odinger equation is no longer integrable because the scattering lengths of the two atomic components are unequal. Even though the soliton is then no longer a soliton in the narrowest sense, we find that the active particle mechanism remains robust as long as a stable solitary wave still exists.

We close in section V with a summary of our results.

\section{II. Ideal Active Soliton Model}

\subsection{II.1 Setup}
For our examination we assume a two-component Bose-Einstein condensate which is confined to one spatial dimension and described by a two-component spinor mean field $\Psi$. The dynamics of such a spinor field $\Psi$, where the atoms have two internal states $\ket{+}$ and $\ket{-}$ and an equal negative inter-component and intra-component scattering length, can be calculated within the mean field approximation using the dimensionless Gross-Pitaevskii equation (GPE) \cite{GPE, BStheo3}
\begin{align} \label{eq:GPE}
i\dot \Psi &= -\frac{1}{2} \Psi^{''} - \braces{\Psi^\dagger \Psi} \Psi + V \Psi + M\Psi + \Gamma \Psi,
\end{align}
where $\Psi = \svek{\Psi_+\von{X,T}}{\Psi_-\von{X,T}}$ is the dimensionless two-component wave function, with $X$ denoting the dimensionless space coordinate, $T$ the dimensionless time, and $V=V\von{X}$ a space dependent external potential. Here and in the following we will denote (where possible) dimensionless quantities by uppercase letters, and their counterpart with dimension by the corresponding lowercase letter. The matrices $M$ and $\Gamma$ are defined by
\begin{align}
M &= \matr{M_+}{0}{0}{M_-}, & \Gamma = \matr{0}{\Gamma_+}{\Gamma_-}{0}.
\end{align}
$M$ can model different chemical potentials $M_\pm$, which can result from different magnetic energies or different trapping strengths of the $\ket{\pm}$ species. $\Gamma$ models a Rabi coupling between both components.  

In adopting this simple form of the dimensionless GPE we have assumed that all one-dimensional interaction constants are negative and equal. For some species of condensate atoms this is naturally true to a good approximation, but if the three-dimensional scattering lengths are unequal then adjustment can be made by tuning the transverse confinement strengths for the different species differently. We will use this dimensionless form of the Gross-Pitaevskii equation \eqref{eq:GPE} throughout section II, but we pause briefly here to relate it to physical units. 

\subsubsection{(a) Physical units}
For repulsively interacting condensates one typically introduces dimensionless variables based on the so-called \emph{healing length} which is proportional to the square root of the interaction constant times a typical density scale, but for attractive interactions this is awkward; while we can easily take the absolute value of the interaction constant, there is no ambient density scale because an attractively interacting condensate is unstable to breaking up into bright soliton droplets. For three-dimensional scattering length $a$, radial trapping frequency $\omega_\rho$, atom mass $m$ and total atom number $N$, however, we can always define a characteristic velocity
\begin{align} \label{eq:v0}
v_0 = \frac{2N}{\pi} \omega_\rho a,
\end{align}
and measure time, space and the density of the wave function in units of
\begin{align}
t_0 &= \frac{\hbar}{m} \frac{1}{v_0^2}, &
x_0 &= \frac{\hbar}{m} \frac{1}{v_0}, &
\psi_0 &= \sqrt{\frac{m}{\hbar} \frac{v_0}{n}},
\end{align}
to achieve a description in dimensionless variables. For the experimental data of \cite{Khaykovich}, for example, we obtain $v_0 \approx 3.6 \text{ mm/s}$, i.e. $t_0 \approx 0.12 \text{ ms}$ and $x_0 \approx 0.43 \text{ } \mathrm{\upmu m}$.

\subsubsection{(b) Coupling terms}
Without external potential and Rabi coupling, i.e. $V=\Gamma=0$, equation \eqref{eq:GPE} has solutions in form of vector bright solitons. The fact that these hyperbolic secant solutions are exact without $V$ and $\Gamma$ is not as important to us as the fact that these soliton forms remain excellent approximate solutions in the case where there are non-vanishing external and coupling potentials \cite{RabiSwitch}. Without loss of generality we will choose in the following
\begin{align} \label{eq:M}
M &= \frac{1}{2}\Omega\matr{1}{0}{0}{-1},
\end{align}
with the real, positive parameter $\Omega$ describing the energy gap between the two chemical potentials of the two components of the condensate. To achieve an energy transfer from the internal to the mechanical degrees of freedom, we choose the coupling
\begin{align}
\Gamma &= \Gamma_0 \matr{0}{\frac{1}{2}\exp\von{iKX}}{\frac{1}{2}\exp\von{-iKX}}{0}, \label{eq:Gamma}
\end{align}
with the real, positive parameter $K$ describing the spatial period of the Rabi coupling with strength $\Gamma_0$ and the external potential
\begin{align} \label{eq:V}
V\von{X}=GX,
\end{align}
with the real, positive parameter $G$ describing the strength of the linear potential against which the soliton is to be lifted.

\subsection{II.2 Numerical results}
We solve \eqref{eq:GPE} numerically, starting with a bright soliton which has an almost vanishing second component and an initial speed $v=1.2 v_0$:
\begin{align} \notag
\Psi_+\von{X,T=0} &= \frac{1}{2} \sqrt{0.998} \sech{\frac{1}{2}X} \exp\von{1.2 iX} \\ \label{eq:initfield2}
\Psi_-\von{X,T=0} &= \frac{1}{2} \sqrt{0.002} \sech{\frac{1}{2}X} \exp\von{1.2 iX}.
\end{align}
It is important that this initial speed is greater than the critical speed $v_\text{c} = \frac{\Omega}{K}v_0$  (see section III.2), but any other initial speeds above $v_\text{c}$ would produce qualitatively similar evolution. We take $K=\Omega=1$, $\Gamma_0 = \pi / \braces{800\csch{\pi}}$ and $G = 1/1600$ as system parameters; these precise values have no qualitative significance for the evolution, but their orders of magnitude do (again see section III.2).

\begin{figure}[thb]
	\centering
	\includegraphics*[trim=162 213 190 230 , width=0.45\textwidth]{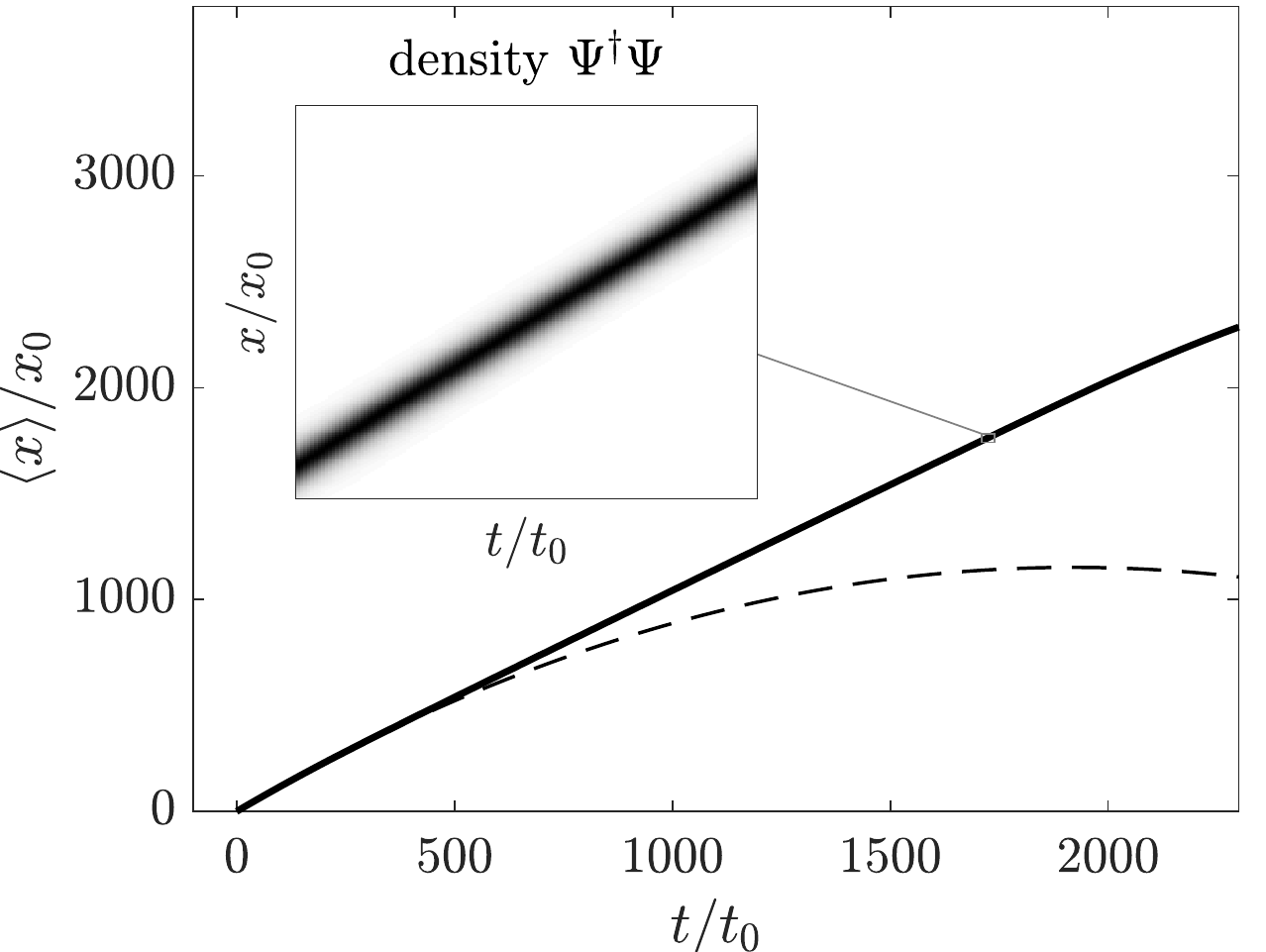}
	\caption{\label{fig:demo} The expectation value $x/x_0$ of the soliton plotted over time for the parameters given in the text (solid), and for the uncoupled case (dashed). While the uncoupled case shows a ballistic curve the velocity stays about constant if the coupling is apparent. The inset shows the density of the solitonic field around $t/t_0=1700$.}
\end{figure}
The behavior of the soliton can be concluded from Fig.~\ref{fig:demo}, where the expectation value of the space coordinate 
\begin{align}
\langle X \rangle = \infinteg{X} \Psi^\dagger X \Psi
\end{align}
is plotted over time $T$ for the former described parameters and initial fields as solid, black line. The uncouple case with the same parameters, except $\Gamma_0 = 0$, is shown as dashed line. While in the uncoupled case the soliton shows a ballistic curve, the position of the soliton increases linearly in time with coupling. Keeping a constant speed greater zero means also that the soliton keeps its kinetic energy, while increasing its potential energy linearly. Since the energy of the total system is constant, this energy can only originate from the internal Zeeman energy. The inset in Fig.~\ref{fig:demo} shows the density $\Psi^\dagger \Psi$ at $T \approx 1700$ for about a range of $40$ units of $X$ and $30$ units of $T$. There are no changes in the density notable.

To demonstrate that the soliton indeed shows the dramatic behavior of transferring energy from the internal degrees of freedom into motional work, we directly examine the time evolution of the different terms in the system's energy. The Gross-Pitaevskii equation \eqref{eq:GPE} corresponds to the Hamiltonian density
\begin{align} \label{eq:Hdens}
\mathcal{H} = \frac{1}{2}\braces{\Psi^\dagger}'\Psi' - \frac{1}{2}\abs{\Psi^\dagger \Psi}^2 + \Psi^\dagger V \Psi + \Psi^\dagger \braces{M + \Gamma} \Psi.
\end{align}
Within this total energy we can identify the various kinetic, potential, internal (`fuel'), nonlinear and Rabi coupling terms:
\begin{align}
H_\text{kin} &= \frac{1}{2}\infintx \braces{\Psi^\dagger}'\Psi', \label{eq:Ekin}\\
H_\text{pot} &= \infintx \Psi^\dagger V \Psi, \label{eq:Epot}\\
H_\text{fuel} &= \infintx \Psi^\dagger M \Psi, \label{eq:Efuel}\\
H_\text{nl} &= -\frac{1}{2}\infintx \abs{\Psi^\dagger \Psi}^2, \label{eq:Enl}\\
H_\Gamma &= \infintx \Psi^\dagger \Gamma \Psi, \label{eq:Egamma}
\end{align}
and denote the corresponding physical energies as $E_x = \hbar\omega H_x$ for $\omega = \Omega/t_0$. We define further the total mechanical energy of the soliton's motional degree of freedom as
\begin{align} \label{eq:Emech}
E_\text{mech} = E_\text{kin} + E_\text{pot}.
\end{align}
\begin{figure*}[htb]
	\centering
	\includegraphics*[trim=75 235 100 243 , width=0.95\textwidth]{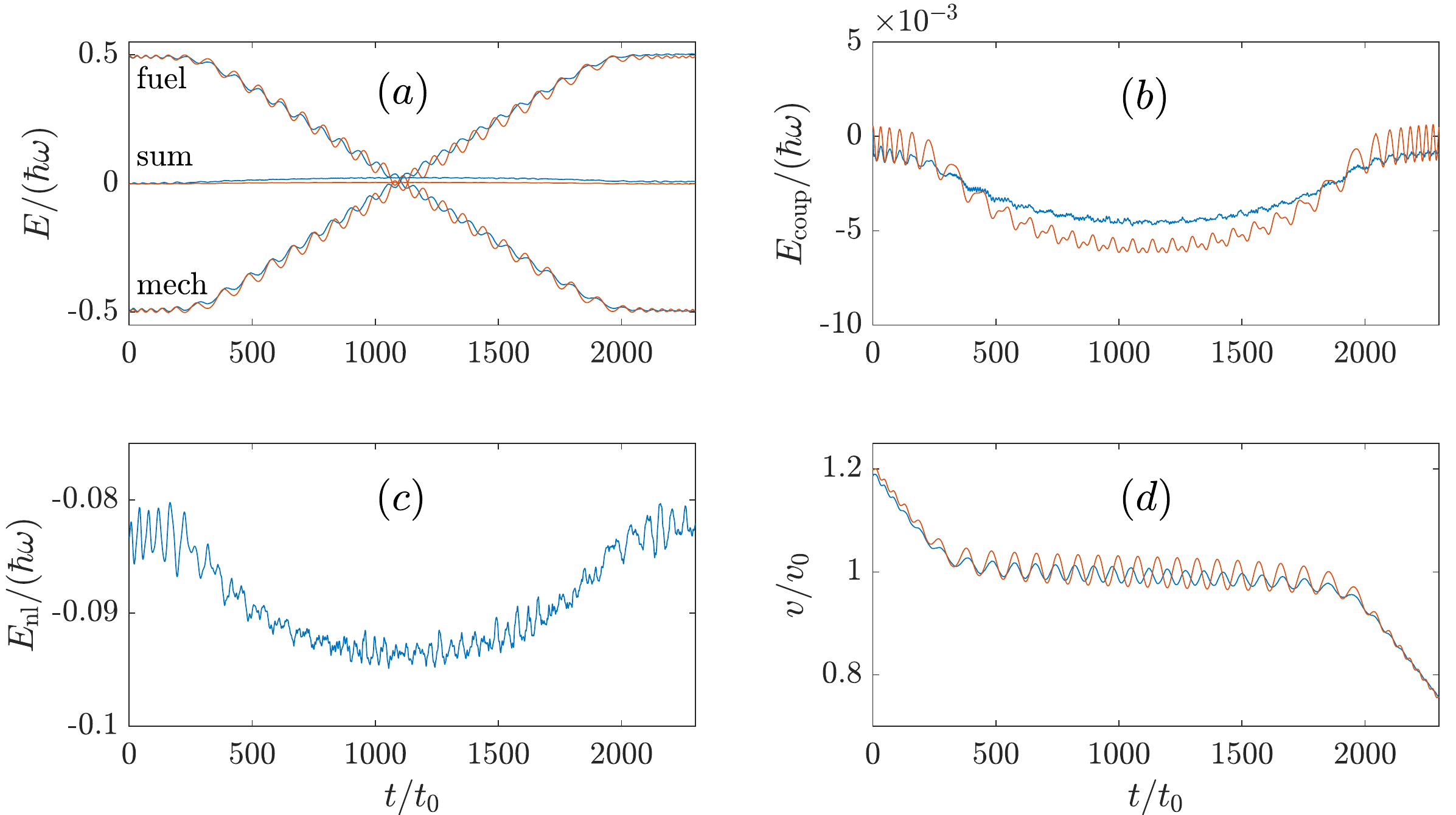}
	\caption{\label{fig:energies} Evolution of the different energy terms of the two-component mean field system (blue) and the collective coordinate model (red, see section III), and the velocity $v/v_0$. (a) The energy in the fuel term decreases linearly during the running phase while the mechanical energy increases linearly, such that the sum of both is about constant. The change of the energy in the coupling term (b) and in the non-linear term (c) are two orders of magnitude below the former energy changes---negligible in comparison. The velocity $v/v_0$ (d) shows oscillations around a constant value during the running phase. Outside the running phase it decreases linearly.}
\end{figure*}

Energies corresponding to \eqref{eq:Ekin}-\eqref{eq:Egamma} are plotted in Fig.~\ref{fig:energies} over time (blue curves). It can be seen in (a) that the total change in the fuel energy $E_\text{fuel}$ and in the mechanical energy $E_\text{mech}$ is of order $\hbar \omega$, and that the sum of both is approximately constant. (We have shifted the origin of $E_\text{mech}$ for convenience.) The maximal changes in the coupling energy $E_\Gamma$ (b) and the energy of the non-linear term $E_\text{nl}$ (c) are in contrast only of order $0.01\hbar\omega$, insignificant compared to the energy transfer between $E_\text{pot}$ and $E_\text{fuel}$. What the couplings are therefore doing is sustaining this steady energy transfer from $E_\text{fuel}$ to $E_\text{pot}$, rather than directly contributing their own energy to the motion. The velocity of the soliton can be calculated as
\begin{align} \label{eq:v_soliton}
v/v_0 = \infintx \Psi^\dagger \Psi^\prime,
\end{align}
and is shown in Fig.~\ref{fig:energies}(d). During the time at which the energy transfer happens the velocity stays about constant. Outside this time the velocity decreases linearly, as one expects from a particle in a linear potential.


\subsection{II.3 Persistence of solitary wave form}
The one-dimensional nonlinear Schr\"odinger equation with $V=\Gamma=0$ is integrable and the sech-form of our initial state \eqref{eq:initfield2} is a \emph{bright soliton} solution which maintains its spatial profile exactly as it moves at arbitrary speed. With the addition of our Rabi coupling term $H_\Gamma$ and the potential gradient, the system is (presumably) no longer integrable. Our initial state nonetheless evolves as a stable solitary wave, very close to its initial form. We will therefore continue to refer to this moving structure as `the soliton', for the sake of brevity.
\begin{figure}[htb]
	\centering
	\includegraphics*[trim=110 300 140 320 , width=0.45\textwidth]{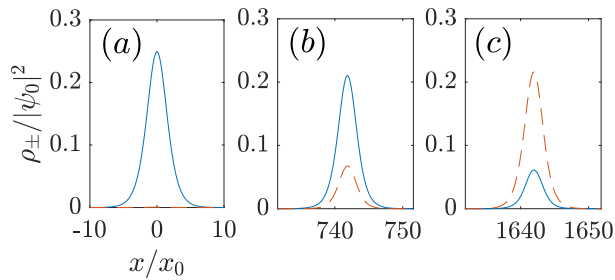}
	\caption{\label{fig:demoplot} The density for the first (second) component plotted in blue (red), for the three different times $T=0$ (a), $T=700$ (b) and $T=1600$ (c). It can be seen that the amplitude of the first component decreases while the amplitude of the second increases which leads to a lower fuel energy. The shape stays the same during the evolution.}
\end{figure}

Figure~\ref{fig:demoplot} shows snapshots of the density $\rho_\pm = \abs{\Psi_\pm}^2$ of the first (second) component of the soliton in blue (red) for the times $T=0$ (a), $T=700$ (b) and $T=1600$ (c). As mentioned above, the initial state has very small density $\rho_-$ in the second component. It can be seen that during the process the density $\rho_+$ decreases while $\rho_-$ increases, which leads to the decrease of the fuel energy $E_\text{fuel}$; this is the energy which lifts the soliton against the external force. It can also be seen from Fig.~\ref{fig:demoplot} that both density profiles remain qualitatively quite similar (though changing in relative size) over the whole evolution, during which the soliton moves many times its own width. How exact can this statement be made?


The general form of a bright soliton for our GPE with $V,\Gamma\to0$ is
\begin{align} \label{eq:freeBS}
\Psi_\pm^\text{bs} \von{X,T} = \frac{1}{2} \sqrt{N_\pm} \sech{\alpha_\pm\braces{X-Q}} e^{i\Phi_\pm},
\end{align}
where $Q$ depends linearly on time, $\Phi_\pm$ depend on time and space, $N_\pm \in \ebraces{0,1}$ (such that $N_+ + N_-=1$) are constant parameters, and $\alpha_\pm=\frac{1}{2}$. From the evolution that we have just seen in Fig.~\ref{fig:demoplot}, it is clear that $N_\pm$ do not remain constant once $\Gamma,V\not=0$. The small but finite change in the non-linear energy $E_\text{nl}$ seen in Fig.~\ref{fig:energies}(c) suggests that $\alpha_\pm$ may not be exactly constant, either. The question is how far the time-dependent $\Psi_\pm$ can still be closely approximated by \emph{some} $\Psi_\pm^\text{bs} \von{X,T}$, for some time-dependent set of parameters. We will find that a close approximation is indeed possible within this bright soliton family, allowing us to pursue a variational model in section III.

To show this we fit the densities $f_\pm=|\Psi_\pm^\text{bs}|^2$ to the numerically obtained densities, using $N_\pm$, $Q$, and $\alpha_\pm$ as tunable fit parameters. As a measure for the deviation of the exactly evolving solution from the soliton form, we define the error $\delta = \delta\von{T}$ between the density $\rho_\pm=\abs{\Psi_\pm\von{X,T}}^2$ and the fitted $f_\pm=\abs{\Psi_\pm^\text{bs}\von{X,T}}^2$ as
\begin{align} \label{eq:function_error}
\delta\von{T} = \sqrt{\infintx \braces{\abs{\rho_+ - f_+}^2 + \abs{\rho_- - f_-}^2}}.
\end{align}
We determine our time-dependent fit parameters as those which minimize this error $\delta\von{T}$ at each instant $T$. 
\begin{figure}[htb]
	\centering
	\includegraphics*[trim=105 295 143 322 , width=0.45\textwidth]{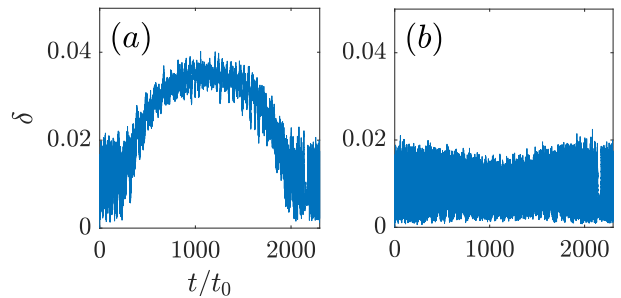}
	\caption{\label{fig:deviation_field_bs} Deviation over time between the numerically exact densities and the soliton ansatz densities for \eqref{eq:freeBS} with instantaneous best-fit parameters. Panel (a) shows the error when $\alpha_\pm$ are not fitted but held at $\alpha_\pm=1/2$; panel (b) shows the error when $\alpha_\pm$ are fitted as well.}
\end{figure}

\begin{figure}[thb]
	\centering
	\includegraphics*[trim=115 195 135 220, width=0.45\textwidth]{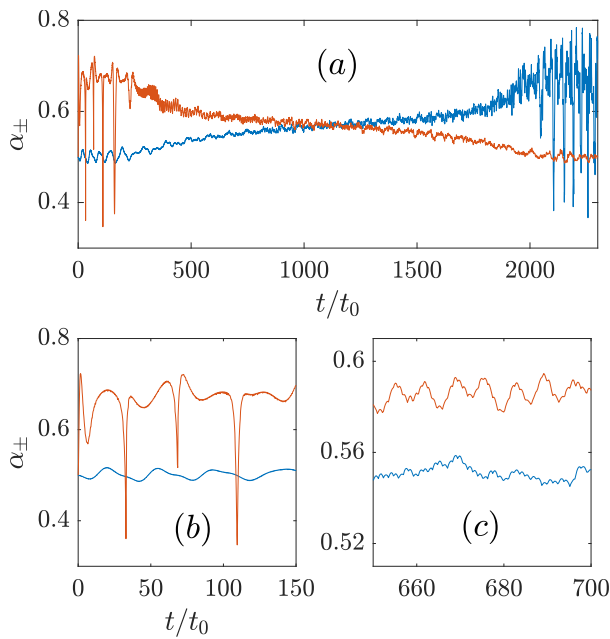}
	\caption{\label{fig:alphafromfit} Width parameters $\alpha_+$ ($\alpha_-$) plotted in blue (red) over time. The total evolution is shown in (a), while (b) and (c) are enlargements at early and at intermediate times. In general the $+$ component in the soliton slowly becomes narrower during the evolution, and the $-$ component broadens after a quick early narrowing; this occurs while the total atomic population also migrates from the $+$ spin component to the $-$, but it is not what we would expect for a vector bright soliton without $\Gamma$ or $G$, since with equal scattering lengths the width of a bright soliton depends only on the total number $N$, not the distribution in spin. When the density of a component is low, large variations in the corresponding width are visible (b). Otherwise the widths show rapid oscillations with a small amplitude (c).}
\end{figure}

Figure \ref{fig:deviation_field_bs}(b) shows $\delta$ plotted over time. The deviation between the exact and fitted-soliton densities remains small, confirming that the densities keep close to the form \eqref{eq:freeBS}. Figure~\ref{fig:deviation_field_bs}(a) shows that the error is not worsened much by leaving out $\alpha_\pm$ as fit parameters and simply keeping $\alpha_\pm = \frac{1}{2}$. As shown in  Fig.~\ref{fig:alphafromfit}, the properly fitted $\alpha_\pm(T)$ do show some understandable trends over longer times, as well as fluctuations on shorter time scales. Substantial departures of $\alpha_\pm$ from $1/2$ only occur at the earliest and latest times, however, when the corresponding $\rho_\pm$ is small. 

The error involved in replacing $\alpha_\pm$ with $1/2$ does not seem, therefore, to represent a significant distortion of the overall $\Psi_\pm$. In section III we will accordingly pursue a time-dependent variational description of our soliton with $\alpha_\pm=1/2$; the expected gain in accuracy from including $\alpha_\pm$ as a variational parameter does not appear to be worth the increased complexity. 

\section{III. Variational approximation for the active bright soliton}

\subsection{III.1 Evolution of the parameters}

Our active solitons remain bright vector solitons to a good approximation, and it is well known that bright solitons behave very much like particles \cite{BStheo1, BStheo2}. This makes it an obvious approach to try to describe the motion of our active soliton more simply in terms of a few collective coordinates, by means of a time-dependent variational approximation. This variational approach is a standard technique for the treatment of solitons \cite{RabiSwitch, ResonantControl, Var1, Var2, Var3, Var4, Var5, Var6, Var7}. 

We begin with the variational Ansatz discussed in section II above:
\begin{align} \label{eq:bs_ansatz}
\Psi^\text{bs}_\pm\von{X,T} = \frac{1}{2} \sqrt{N_\pm} \sech{\frac{1}{2}\braces{X-Q}} \\ \notag
\cdot e^{-iP^2 T} e^{iP\braces{X-Q}} e^{i\Phi_\pm}\;,
\end{align}
where now we allow the Ansatz parameters $Q\von{T},P\von{T},N_\pm\von{T}$ and $\Phi_\pm\von{T}$ to be time-dependent. We then insert this Ansatz into the Lagrangian for the Gross-Pitaevskii equation, to produce a Lagrangian that depends only on our Ansatz parameters (and their time derivatives):
\begin{align} \label{eq:Lagrangian}
L&\von{Q,P,N_\pm,\Phi_\pm} \\ \notag
&= \infintx \sbraces{\mathrm{Im}\braces{\dot \Psi_\text{bs}^\dagger \Psi_\text{bs}} - \mathcal{H}\von{\Psi_\text{bs},\Psi'_\text{bs}}},
\end{align}
with $\Psi_\text{bs} = \svek{\Psi_+^\text{bs}}{\Psi_-^\text{bs}}$ from \eqref{eq:bs_ansatz}. 

Noting that $N_+ + N_- = 1$ is exactly conserved under the GPE, and that the canonically conjugate common global phase of $\Psi_\pm$ is a trivial cyclic variable, we can reduce the number of our parameters by defining
\begin{align} \label{def:canvar}
\Delta &= \Phi_+ - \Phi_-, & I&=\frac{1}{2}\braces{N_+ - N_-}.
\end{align}
This leaves us with two mechanical coordinates $\braces{Q,P}$ to be the position and momentum of the motional degree of freedom of the whole soliton, and $\braces{\Delta,I}$ as internal coordinates describing its total spin degree of freedom as an active particle's internal energy depot. In particular $I$ is directly proportional to the fuel energy $E_\text{fuel}=\hbar\omega I$; $I=1/2$ represents a fully stocked energy depot while at $I=-1/2$ the depot is empty.

With these definitions the Euler-Lagrange equations of motion for our variational Lagrangian read
\begin{align}
\dot Q &= P, \\
\dot P &= -G + K\Gamma_K \sqrt{\frac{1}{4} - I^2} \sin\von{KQ-\Delta}, \label{eq:eom_p} \\
\dot \Delta &= \Omega - \Gamma_K \frac{I}{\sqrt{\frac{1}{4} - I^2}} \cos\von{KQ-\Delta}, \\
\dot I &= -\Gamma_K \sqrt{\frac{1}{4} - I^2} \sin\von{KQ-\Delta},
\end{align}
with the one additional definition, from the Fourier transform
\begin{equation}
\int_{-\infty}^\infty\! dz\,\frac{e^{ikz}}{\cosh^2(z/2)} = \frac{4\pi k}{\sinh\pi k}\;,
\end{equation}
of
\begin{align}
\Gamma_K &= \Gamma_0 \pi K \csch{\pi K},
\end{align}
whereby $\Gamma_K \leq \Gamma_0$ for all $K$. These equations of motion can also be identified as the canonical equations of motion for the Hamiltonian
\begin{align} \label{eq:Hpoint}
H = \frac{1}{2} P^2 + GQ + \Omega I + \Gamma_K \sqrt{\frac{1}{4} - I^2}  \cos\von{KQ - \Delta},
\end{align}
if we assume $\braces{Q,P}$ and $\braces{\Delta,I}$ to be pairs of conjugate coordinates. The several energy terms in  \eqref{eq:Hpoint} are therefore the energies that were plotted in red in Fig.~\ref{fig:energies} for comparison with  the Gross-Pitaevskii energies \eqref{eq:Ekin}-\eqref{eq:Efuel}, \eqref{eq:Egamma} that were plotted in blue. As we saw in Fig.~\ref{fig:energies}(a), the global evolution of the fuel energy $E_\text{fuel}$ and mechanical energy $E_\text{mech}$ is well reproduced by the reduced variational model. Also the variational parameter $P\von{t}$, shown as red line in Fig.~\ref{fig:energies}(d), approximates the velocity of the soliton.  We can therefore seek to understand the behavior of the active bright soliton in terms of this much simpler effective Hamiltonian \eqref{eq:Hpoint}, which has only two degrees of freedom instead of the continuum of the Gross-Pitaevskii order parameter fields. 

The inital state \eqref{eq:initfield2} for the GPE evolution that we showed in section II above corresponds to the initial parameters
\begin{align}
Q(0) &= 0, & P(0) &= 1.2, \nonumber \\
\Delta(0) &= 0, & I(0) &= 0.498
\end{align}
in this reduced representation.

\subsection{III.2 The `Hamiltonian daemon'}


The Hamiltonian \eqref{eq:Hpoint} has in fact been studied at some length already \cite{Daemon1, Daemon2, Daemon4}; it has been labelled a `Hamiltonian daemon' in allusion to Maxwell's Demon and to small autonomous Unix processes (daemons). Indeed the present paper was motivated largely as a deliberate attempt to embed this Hamiltonian daemon into a Gross-Pitaevskii system. We refer readers to  \cite{Daemon1, Daemon2, Daemon4} for a fuller discussion of the Hamiltonian itself, but here we briefly review some of its properties that are important for the behavior of the active soliton.

\subsubsection{(a) The Chirikov resonance}
First of all we can understand the Chirikov resonance that allows secular energy transfer from high frequency into steady motion. The reason why we need a Chirikov resonance is that generically $\Delta$ has a rapid linear dependence on the dimensionless time $T$, $\Delta\sim\Omega T$ with $\Omega$ large because it is the Zeeman frequency in dimensionless form. This generically makes the $\cos(KQ-\Delta)$ term in the Hamiltonian (\ref{eq:Hpoint}) oscillate rapidly, and thus time-average to zero so that the fuel and motional degrees of freedom adiabatically decouple. 

The Chirikov resonance occurs, however, if the soliton moves at nearly the (dimensionless) speed $P_c = \Omega/K$. At this special speed we have $Q\sim \Omega T/K$, so that $\cos(KQ-\Delta)$ no longer oscillates rapidly around zero but becomes nearly constant instead. The Rabi coupling term, which is represented in the variational $H$ as the $\Gamma_K \sqrt{\frac{1}{4} - I^2}  \cos\von{KQ - \Delta}$ term, can therefore have a secular effect. 

What is not generic for Chirikov resonances, but occurs in this particular `daemon' case, is that the secular effect of this interaction term which appears at the Chirikov resonance $P =P_c$ is to \emph{keep} $P$ close to $P_c$, so that the resonance actually sustains itself, allowing secular energy transfer to continue for a long time. If the (dimensionless) wave number $K$ of the Rabi coupling's spatial periodicity is large enough, $K$ can effectively `gear down' a high frequency $\Omega$ to an arbitrarily slow steady speed $P_c = \Omega/K$. 

\subsubsection{(b) Parameter regimes}
As well as understanding how the active soliton can basically work, we can identify the parameter regimes in which the soliton's nontrivially driven motion represents a reasonable form of active particle. First of all we can see that the soliton can only be active if 
\begin{equation}
G<K\Gamma_K/2\;,
\end{equation} 
since otherwise $\dot{P}$ will always be negative according to (\ref{eq:eom_p}), and therefore cannot remain near the Chirikov resonance at $P=P_c$ for any long time. In effect $K\Gamma_K/2$ represents the largest force which the active particle's `motor' can exert; an external downward force stronger than this limit will prevent upward motion, in the same way that a car with limited engine torque cannot climb too steep a hill. 

A second condition for an active particle is what we were able to achieve in section II: steady motion over a long time because a mechanism is slowly transferring a large amount of energy from fuel into work, while the energy of the mechanism itself remains comparatively small at all times. This requires the regime 
\begin{equation}
\Gamma_K \ll \Omega\;.
\end{equation}

Finally, for small $\Gamma_K$ the Chirikov resonance is narrow and can only sustain itself once $P$ is quite close to $P_c$. The active soliton must therefore somehow be given the initial kinetic energy $P_c^2/2$ before the depot energy $\Omega I$ can be exploited for motion. To make this initial energy investment worthwhile, an active soliton should be able to hold at least that much energy in its depot, requiring
\begin{equation}
P_c^2 < 2\Omega 
\end{equation}
and hence
\begin{equation}
\Omega < 2K^2\;.
\end{equation}

Since any one of our four dimensionless parameters $G, \Gamma_K, \Omega$ and $K$ can effectively be set to one by rescaling the dimensionless time $T$, the inequality hierarchy
\begin{equation}\label{regime}
\frac{2G}{K} < \Gamma_K \ll \Omega < 2K^2
\end{equation}
fully defines the regime in which this daemon-soliton can reasonably considered as an active particle. As noted in \cite{Daemon1, Daemon2, Daemon4}, the strong inequality which is the middle condition in (\ref{regime}) implies a time scale hierarchy in the evolution under (\ref{eq:Hpoint}) which allows adiabatic methods to be applied, clarifying some otherwise rather complicated nonlinear dynamics.
For reference, parameters used in section II satisfy the hierarchy (\ref{regime}) with
\begin{equation}
\frac{1}{800} <  \frac{\pi^2}{800} \ll 1 < 2\;.
\end{equation}

\subsubsection{(c) Performance limits}
The total time $T_\text{c}$ during which the soliton can remain active before all of its fuel has been expended ($I$ falls from $+1/2$ to $-1/2$) can be estimated as $\Omega$ (maximum total fuel energy) divided by power needed to sustain speed $P_\text{c} = \Omega/K$ against external force $G$. This yields 
\begin{align} \label{eq:typicalParameters}
T_\text{c} &= \frac{K}{G}, &
X_\text{c} &= P_\text{c} T_\text{c} = \frac{\Omega}{G}\;.
\end{align}
If we use the data from \cite{Khaykovich} to translate our parameters from section II into physical units, we find a critical velocity $P_\text{c} v_0 \approx 3.6 \text{ mm}/\text{s}$, active motion duration $T_\text{c} t_0 \approx 0.2 \text{ s}$, and total height raised $X_\text{c} x_0 \approx 0.7 \text{ mm}$. These appear not unreasonable as experimental dimensions; they fall well short of the potential maximum height to which Zeeman energy might carry a bright vector soliton, but with soliton widths in microns, a million times the soliton width might require an extravagantly large vacuum chamber.

\subsubsection{(d) Daemon `ignition'}
\begin{figure}[thb]
	\centering
	\includegraphics*[trim=105 195 140 220 , width=0.45\textwidth]{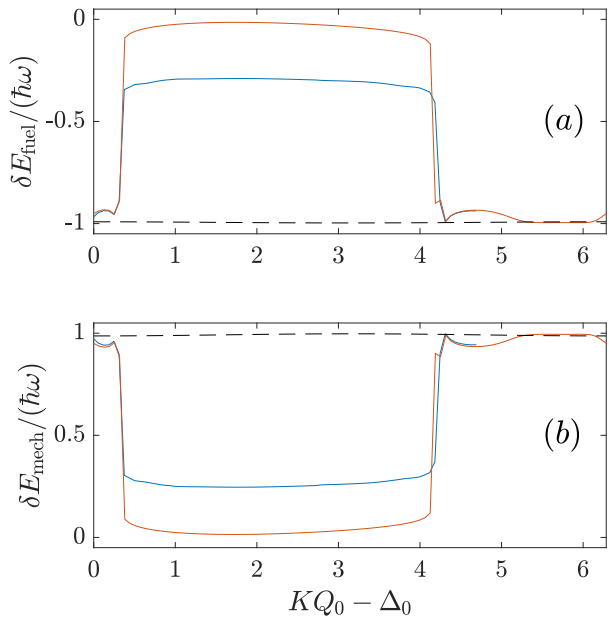}
	\caption{\label{fig:ignition} (a) The relative change in the fuel energy and (b) in the mechanical energy during the process plotted over the initial angle $K Q_0 - \Delta_0$. For a weak coupling $\Gamma_K$ two regions can be identified for the particle system (red) and the field system (blue). In the one region almost all energy of the fuel gets transferred to the mechanical degrees of freedom of the soliton, in the other region much less energy gets transferred. This effect does not occur for a stronger coupling (dashed black line, both cases), for which complete energy transfer occurs for all initial angles.}
\end{figure}

Because the evolution of the Hamiltonian daemon is indeed quite complicated, the accuracy of the collective coordinate model in representing the active soliton can be limited in some subtle ways. 
For example in Fig.~\ref{fig:ignition} we show two cases with $K=\Omega=1$ and $G=\pi^2/1600$ as in section II, but with either $\Gamma_K = \pi^2/800$ as in section II or $\Gamma_K = \pi^2/3200$, four times smaller than in section II. Figure~\ref{fig:ignition}(a) shows the relative change of the fuel energy $\delta E_\text{fuel} / \braces{\hbar\omega}$ between the start of the process, and after an evolution of time $\delta T=2300$ for different initial angle $KQ_0-\Delta_0$. Analogously Fig.~\ref{fig:ignition}(b) show the relative change in the mechanical energy. The small-$\Gamma_K$ cases are shown as solid lines, blue for the field system and red for the variational system. The big-$\Gamma_K$ cases are shown as dashed black lines. The field and variational calculation yield only not notable deviations from each other in this case. While the dashed big-$\Gamma_K$ solutions show the same behavior as could be seen before (the energy is transferred from $H_\text{fuel}$ to $H_\text{mech}$), the solutions for the smaller $\Gamma_K$ do not show this behavior on the whole range of initial values.

With smaller $\Gamma_K$ the Chirikov resonance is narrower, and the system does not always get captured into the resonant region of phase space~\cite{Liouville}. Whether or not active motion occurs depends in this small-$\Gamma_K$ case on the initial value of the angle $\Delta_0$. This phenomenon of state-dependent `ignition' of the active phase occurs qualitatively in both the reduced variational model and the full Gross-Pitaevskii description, but as Fig.~\ref{fig:ignition} shows the non-ignition effect is different, when it occurs, in the full theory and in the variational approximation.

When not all of the spin energy is successfully consumed for motional work, the variational approximation predicts that very little of it will be used in any way---the `motor' simply fails to start. In these cases where the variational approximation predicts little change in spin energy, however, the full GPE evolution shows some significant loss of $E_\text{fuel}$. Closer examination shows that what is occurring in these cases is that correspondence between the field theory and the variational model breaks down, because passing through the Chirikov resonance makes the soliton partly break up; spin energy is used to unbind part of the soliton, rather than to keep the whole soliton moving against the external force. It is perhaps counter-intuitive that this happens for \emph{weaker} $\Gamma_K$ coupling, while the stronger coupling keeps the soliton together reliably, as well as reliably using its spin energy as an active particle. The reason for this is that the bigger coupling forces the system to get caught in the Chirikov resonance. If the system is not caught in the resonance, which can only happen for the small-$\Gamma_K$ case, the effect of the Rabi coupling is rather that it transfers a part of the soliton from the + component to the - component of the condensate, while increasing the momentum only of the transferred part, which will hence leave the soliton.


This example shows that although our simple picture of the vector bright soliton as an active particle certainly can apply well, with the simple daemon model as an accurate variational approximation, yet there can also be cases where the soliton fails to remain robust under the spatially periodic Rabi coupling and the external force. Interesting as it may be to pursue a detailed investigation of exactly what happens in the particular case of Fig.~\ref{fig:ignition}, we will leave this for future work; we conclude the present paper by looking more broadly at how our active particle model may break down, or survive, when the condensate system is generalized in various ways.

\section{IV. More general active soliton models}
\subsection{IV.1 Other Rabi couplings}
In \eqref{eq:Gamma} we introduced a Rabi coupling that was tailored specifically to realize the Chirikov resonance of the Hamiltonian daemon for an atomic soliton. To consider engineered Hamiltonians that do not involve Chirikov resonances at all would give this paper an infinite scope, but we can consider a wider range of qualitatively similar Chirikov resonances. In particular in this subsection we modify our assumption that the Rabi coupling is perfectly periodic in space with a single positive wave number $K$. Instead we will examine a case with multiple simultaneous $K\to K_n$; a case with negative $K$, in which the active soliton harvests internal energy from the external force instead of working against it; and a case where the periodicity of the Rabi coupling is somewhat disordered, so that $K$ has a finite statistical width.

All of these cases still have
\begin{align}
\Gamma &= \Gamma_0 \matr{0}{F(X)}{F^*(X)}{0},
\end{align}
for some spatially varying function $F=F\von{X}$. Letting $\tilde F$ denote the Fourier transform of $F$, we can express our spatially dependent Rabi coupling as
\begin{align}
&H_\Gamma \equiv \infintx \Psi_\text{bs}^\dagger \Gamma \Psi_\text{bs} \nonumber \\ \label{eq:Hgamma_allg}
&= \frac{\Gamma_0}{2}\sqrt{\braces{\frac{1}{2}}^2-I^2} \text{Re} \von{ e^{i\Delta} \infinteg{X} {F}\von{X}\mathrm{sech}^2\von{\frac{X-Q}{2}}}\nonumber\\
&= \Gamma_0 \sqrt{\braces{\frac{1}{2}}^2-I^2}\; \text{Re} \von{ e^{i\Delta} \infinteg{Y} \frac{Y\tilde{F}\von{Y}} {\sinh\von{\pi Y}}  e^{-iQY}},
\end{align}
again using the Fourier transform of $\mathrm{sech}^2$.
Where our original ansatz \eqref{eq:Gamma} made $\tilde F$ a delta function at $Y=K$, we now explore some more general cases. We find that the mechanism of energy transfer through the self-sustaining Chirikov resonance of the Hamiltonian daemon does not only occur in our original special case, but persists much more generally. Embedding the Hamiltonian daemon in the spinor condensate mean field does introduce generalisations and corrections to the most basic daemon Hamiltonian, however.

\subsubsection{(a) A coupling with 3 isolated resonances}
Since the complex exponential $F=e^{iKX}$ in our original model explicitly broke time-reversal invariance, one might well worry that the active soliton can only work when this important discrete symmetry is broken by hand. If instead we take $F\von{X} = 2 \cos^2\von{KX/2}$, however, we maintain time-reversal invariance with the real Hamiltonian term
\begin{align}\label{HGam2}
H_\Gamma = \sqrt{\frac{1}{4}-I^2} ( 2\Gamma_0\cos\von{\Delta} + \Gamma_K \cos\von{KQ-\Delta} \notag \\
+ \Gamma_K \cos\von{KQ + \Delta} )\;.
\end{align}
Our previous `$K$' Chirikov resonance is now joined by similar `$-K$' and `$K=0$' resonances: a term in $H_\Gamma$ is now time-independent when $\dot{Q}=P$ is close to $-P_c=-\Omega/K$ or close to zero. The presence of additional well separated resonances turns out to have little effect, however, because with $\Gamma_0\ll \Omega$ each term in $H_\Gamma$ can only have significant effect when $P$ is close to its critical speed. The soliton cannot be close to more than one critical speed simultaneously, because the three are separated by $\Omega/K$, and so effectively this time-reversal-invariant model is nearly equivalent to our original model, as long as the soliton speed is not close to either zero or $-P_c$.

\begin{figure}[thb]
	\centering
	\includegraphics*[trim=105 195 140 225, width=0.45\textwidth]{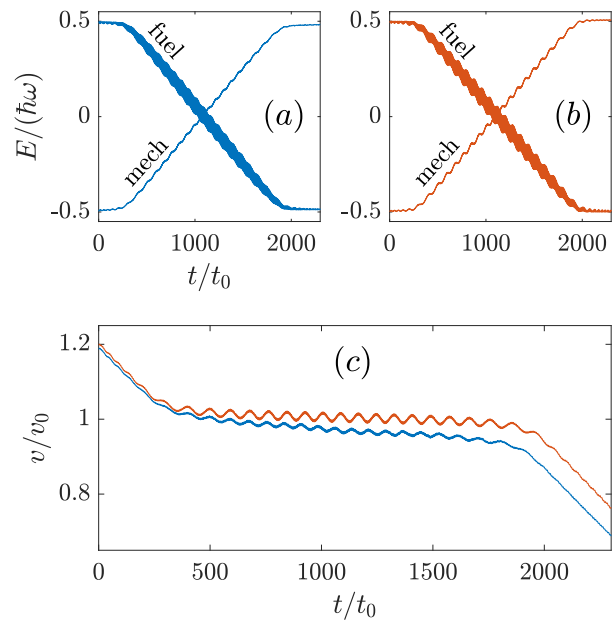}
	\caption{\label{fig:different_res} Evolution of the fuel and mechanical energy ((a) and (b)) and the velocity $v/v_0$(c), for the case of multiple resonances. The blue (red) lines correspond to the field (particle) system. While there are rapid oscillations on top of the global evolution, the global behavior is almost the same as for a single resonance.}
\end{figure}

Figure \ref{fig:different_res} shows numerical evolutions with this coupling (\ref{HGam2}), but otherwise with all parameters and initial conditions the same as in section II. In the Figure both the full GPE field theory (blue) and the corresponding variational approximation (red) are shown. The behaviors of the fuel energy $E_\text{fuel}$, the mechanical energy $E_\text{mech}$, and the velocity $v$ are qualitatively very much the same as for our original single '$K$' resonance: energy is slowly drained from the fuel term and converted into mechanical energy. 

The new terms in the Hamiltonian introduce additional Chirikov resonances; that is, these new terms would be resonant perturbations of the system at speeds zero or $-P_c$. As long as the active soliton has $P$ near $P_c$, the additional terms are far from being resonant. Their effects on the motion of the soliton near $P_c$ are therefore not the dramatic qualitative change of a resonant perturbation, but only `dressing': the non-resonant terms can be adiabatically eliminated and replaced, in an adiabatic effective Hamiltonian with renormalized terms. Such an effect can be seen in Fig.~\ref{fig:different_res}, for example. When the exact soliton velocity $v/v_0$ from \eqref{eq:v_soliton} is compared with the variational parameter $P(T)$, we see that there is a small linear downward trend in the exact velocity which is not captured by the variational model. 
Such a downward trend in the active particle speed, as its fuel $I$ is depleted, can be incorporated into the Hamiltonian daemon by adding a quadratic term $b I^2$ for some $b>0$ to the daemon Hamiltonian \eqref{eq:Hpoint}. Doing that effectively makes $\Omega \to \Omega + 2bI$ and thus $P_c -> P_c + (2b/K)I$, decreasing as $I$ decreases. Within the full mean field theory this effect can easily arise, because the vector bright soliton distorts somewhat in form during the active phase, as we have already seen in the non-trivial behavior of the $\alpha_\pm$ variational parameters shown in Fig.~\ref{fig:alphafromfit}. What evidently occurs with the additional couplings of (\ref{HGam2}) is that the soliton slightly distorts in an $I$-dependent way, such that the nonlinear and kinetic terms in the energy contribute an effective $bI^2$ energy term.

\subsubsection{(b) Working as a brake}
We have just seen that for a more general $H_\Gamma$ the active soliton's speed may be generalized away from the single sharp $P_c$ of the simplest model. We can now study a more drastic speed alternative for the active soliton while still using the same three-resonance $H_\Gamma$  (\ref{HGam2}). With time-reversal symmetry now restored, we can find a case in which the '$-K$' resonance generates the major effect of holding the soliton's speed nearly steady as it moves with the external force instead of against it. In other words the Chirikov mechanism which previously worked as a motor now works as a brake. Since the internal energy $I$ is now raised instead of being depleted, we could also consider this evolution as a microscopically described case of energy harvesting by an active particle.

\begin{figure}[thb]
	\centering
	\includegraphics*[trim=105 195 140 225 , width=0.45\textwidth]{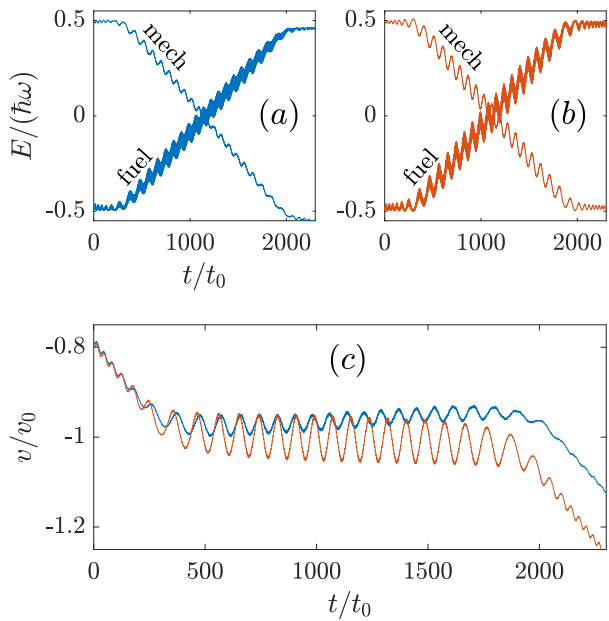}
	\caption{\label{fig:brake} Evolution of the fuel and mechanical energy ((a) and (b)) and the velocity (c) for the case, where the velocity locks to a negative value. The field (particle) system is plotted in blue (red). Instead of getting accelerated the soliton keeps its negative velocity, while the energy of the fuel term is increased.}
\end{figure}

Initial conditions which lead to this motion are
\begin{align} \notag
\Psi_+\von{X,T=0} &= \frac{1}{2} \sqrt{0.002} \sech{\frac{1}{2}X} \exp\von{-0.8 iX} \\ \label{eq:initfieldbrake2}
\Psi_-\von{X,T=0} &= \frac{1}{2} \sqrt{0.998} \sech{\frac{1}{2}X} \exp\von{-0.8 iX}.
\end{align}
These initial fields mean that $I\von{T=0}=-0.498$, \emph{i.e.} there is initially almost no energy in the fuel term, and the soliton starts with the negative velocity $-0.8v_0$, falling down the linear potential. Due to the potential gradient the soliton will linearly increase its momentum, until it reaches $P \approx -1$. It then becomes trapped into the '$-K$' resonance, locking its momentum close to $-P_c=\frac{\Omega}{-K} = -1$. The mechanical energy which is lost as the soliton moves down the potential gradient without correspondingly accelerating is transferred into the internal depot of spin energy. This behavior can be seen in Fig.~\ref{fig:brake}, which can be directly compared to Fig.~\ref{fig:different_res}. 

Much as in Fig.~\ref{fig:different_res}, a small but systematic discrepancy between the variational approximation and the full field theory can be seen in Fig.~\ref{fig:brake}. The active soliton not only avoids accelerating despite the external force, but actually decelerates slightly (in the negative direction, \emph{i.e.} becomes slightly less negative). The same form of $bI^2$ modification to (\ref{eq:Hpoint}) can account for this residual deceleration, but now we will need $b<0$, so that $P_c = (\Omega+2bI)/(-K)$ decreases in absolute value as $I$ rises. Evidently the way in which soliton deformation dresses the effective daemon Hamiltonian depends on which Chirikov coupling is actually resonant.

\subsubsection{(c) Finite-width resonance}
As a last example in our exploration of more general $F(X)$ coupling profiles, we return to the original case with only one Chirikov resonance. We now suppose, however, that the Fourier transform of $F$ is not a sharp delta function at $K$, but has a small finite width $\sigma$, because the Rabi drive does not extend over all space, but instead has a Gaussian envelope with a large width $1/\sigma$:
\begin{align}
F\von{x} = \frac{1}{\sqrt{2\pi}} \exp\von{-iKX} \exp\von{-\frac{1}{2}\sigma^2 X^2}\;.
\end{align}
Since this means that the active motion of the soliton can only occur within the Gaussian envelope, we are interested in $\sigma$ small enough that $1/\sigma$ still allows room for a reasonable amount of active motion. Below we will compare two cases: $\sigma = 0.001$ and $\sigma=0.002$, for both of which $1/\sigma$ is on the order of the total travel distances that we saw in section II.

For such small $\sigma$ the spatial modulation of $F(X)$ is slow on the $1/K$ scale over which the Chirikov resonance motor effect operates, and so we can anticipate behavior much like what we have previously seen for the single-resonance model, except with a $\Gamma_K$ that slowly changes as the soliton moves. Quantitatively, we expect that the active motion can only persist as long as the instantaneous coupling strength $\Gamma_K$ is above the $2G/K$ threshold from (\ref{regime}), so that the maximum possible force from the Rabi coupling can overcome the external force $G$. 

To test these predictions and also continue our comparisons between GPE evolution and the variational approximation with two degrees of freedom, we should ideally define $H_\Gamma$ using the integral in \eqref{eq:Hgamma_allg}. Unfortunately this integral has no convenient analytical form. We therefore instead take a more transparent $H_\Gamma$ with the same Gaussian form as $F(X)$ itself, and allow for modification due to averaging over the soliton width by using fitting parameters $\alpha$ and $\beta$:
\begin{align}
H_\Gamma = \Gamma_K(Q)  \sqrt{\frac{1}{4}-I^2} \cos\von{K_0 Q - \Delta},
\end{align}
where we define
\begin{align} \notag
\Gamma_K(Q) &= \Gamma_0 \frac{\alpha}{\sqrt{1+2\beta\sigma^2}} \exp\von{-\frac{\beta K^2}{1+2\beta\sigma^2}} \cdot \\
&\qquad\qquad\cdot\exp\von{-\frac{\sigma^2 Q^2}{2\braces{1+2\beta\sigma^2}}},  \\
K_0 &= \frac{K}{1+2\beta\sigma^2}\;.
\end{align}
Best fits to the ideal $H_\Gamma$ from the \eqref{eq:Hgamma_allg} integral are for $\alpha \approx 0.969$ and $\beta \approx 1.269$.

\begin{figure}[thb]
	\centering
	\includegraphics*[trim=105 120 130 135 , width=0.45\textwidth]{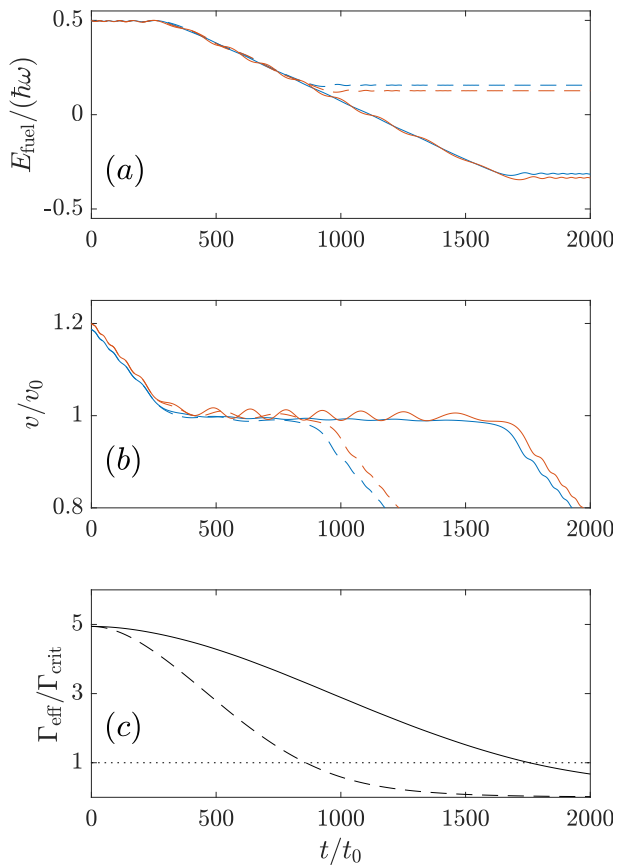}
	\caption{\label{fig:finitefield} Evolution of the fuel energy $E_\text{fuel}/\braces{\hbar \omega}$ (a) and the velocity of the soliton (b) for the case of a finite resonance. The blue (red) curves correspond to the field (particle) system, the solid (dashed) curves to the narrow (wide) resonance width. The begin of the evolution is the same as for a sharp resonance. The process stops when the effective coupling strength $\Gamma_\text{eff}$  drops to the critical value $\Gamma_\text{crit}$ (c) even though the fuel has not yet been drained completely.}
\end{figure}

Figure~\ref{fig:finitefield} shows in (a) and (b) the evolution of the fuel energy and velocity of the soliton according to both the GPE and the variational approximation, for the two different resonance widths $\sigma = 0.002$ (dashed) and $\sigma = 0.001$ (solid). All system parameters and initial values are otherwise the same as for the evolutions in section~II. The soliton thus begins at $X=0$, in the middle of the Rabi coupling's Gaussian envelope, at a speed slightly above the critical speed for active motion $P_c$. Decelerating under the external force, its speed soon falls to $P_c$ and active motion begins. With these two particular resonance widths the now $Q$-dependent coupling $\Gamma_K(Q)$ drops below the critical value $2G/K_0$ before the active motion would otherwise stop for lack of internal energy. 

For $\sigma = 0.002$ this happens at about $T\approx 720$, and the active phase of motion indeed ceases then, with the fuel energy holding constant after this point, and velocity decreasing linearly under the external force. For the narrower resonance width $\sigma = 0.001$, the local coupling constant $\Gamma_K(Q)$ reaches the critical value at $T\approx 1600$, but the active phase ends noticeably earlier, at $T\approx 1400$, when $\Gamma_K(Q)\doteq 2.8 G/K_0$. It must be remembered, however, that the condition $\Gamma_K > 2G/K$ is the requirement for active motion \emph{at any possible} $I$. The force exerted by the Rabi coupling is also proportional to $\sqrt{(1/4)-I^2}$, so active motion for $|I|>0$ in general requires higher $\Gamma_K$. For the spatially narrower Rabi drive envelope with $\sigma=0.002$, the too-small-$\Gamma_K$ threshold is crossed sooner, when $I$ is close to zero, while for the spatially wider case $\sigma=0.001$ the active motion continues until $I$ is near $-0.3$. At that point the $\sqrt{(1/4)-I^2}$ factor and the smaller $\Gamma_K(Q)$ together leave the Rabi drive too weak to overcome the external force, and active motion stops.

Overall our investigations of different forms of spatially dependent Rabi coupling support the view that although the Chirikov resonance mechanism of the active vector bright soliton is somewhat complicated, it is not especially fragile. The many possible perturbations and parameters of the larger field theory model can impose many deformations and renormalizations of the simple Hamiltonian daemon model, which can effect its behavior significantly. They do not destroy its operation entirely, however; on the contrary, the basic active particle behavior can robustly persist.

\subsection{IV.2 Different scattering lengths}

The form \eqref{eq:GPE} of the one-dimensional Gross-Pitaevskii equation does not hold if the three-dimensional inter-species scattering lengths $a_{\pm \pm'}$ and radial trapping frequencies $\omega_{\rho\pm}$, which together determine the effective one-dimensional nonlinear coupling matrix, are not all equal for the two species. While one coupling constant can be set equal to one by scaling, in general the other three are independent and our GPE \eqref{eq:GPE} must be generalized by replacing
\begin{align} \label{eq:nonlin_gen}
\braces{\Psi^\dagger \Psi}\Psi \to
\svek{\braces{\abs{\Psi_+}^2 + A'\abs{\Psi_-}^2}\Psi_+}{\braces{A'\abs{\Psi_+}^2 + A\abs{\Psi_-}^2}\Psi_-}
\end{align}
for some real constants $A$ and $A'$. By modifying the internal states of the atoms with external fields, the three-dimensional scattering lengths can be controlled experimentally to a great extent; by further making the potentials which confine the gas to one dimension spin-dependent, the mean-field coupling constants in one dimension can further be modified. It may not be experimentally easy to achieve $A=A'=1$, however, and so we ask what effects general $A$ and $A'$ may have on the active bright vector soliton.

\subsubsection{(a) Variational Hamiltonian}
The main effect of general $A$ and $A'$ is that the soliton is no longer a soliton in the strict sense, since the general two-component nonlinear Schr\"odinger equation is no longer integrable. Similar solitary wave solutions still exist for a range of $A$ and $A'$, however, and we will continue to refer to these as `the soliton' for brevity. Their wave function profile is no longer exactly a hyperbolic cosecant, and in general the widths of the solitary wave are not the same for both components $\Psi_\pm$. For simplicity, however, we will continue to compare numerical GPE evolution with a variational approximation based on the same ansatz (\ref{eq:bs_ansatz}) that we used before, with hyperbolic cosecants of the same fixed width. 

This simpler variational ansatz leaves most of our effective variational Hamiltonian (\ref{eq:Hpoint}) unaffected by the new set of scattering lengths, even though a more accurate ansatz would include indirect effects due to the modification of the solitary wave shapes. Even without taking those changes into account, however, additional $I$ dependences appear in the variational $H$ through the nonlinear term in the Gross-Pitaevskii Hamiltonian density, in which we must replace $\abs{\Psi^\dagger \Psi}^2$ in (\ref{eq:Hdens}) with
\begin{align} \label{eq:replaceG}
&\frac{1}{2}\abs{\Psi^\dagger \Psi}^2 \\
&\to
\frac{1}{2} \braces{\abs{\Psi_+^* \Psi_+}^2 + A\abs{\Psi_-^* \Psi_-}^2 + 2 A' \abs{\Psi_+}^2 \abs{\Psi_-}^2}\;. \notag
\end{align}
Using the integral
\begin{equation}
\int_{-\infty}^\infty\!dz\,\mathrm{sech}^4(z) \equiv \int_{-\infty}^\infty\!d(\tanh z)\,(1-\tanh^2z) = \frac{4}{3}
\end{equation}
we see that we now have
\begin{align} \label{eq:Hpoint2}
H = &\frac{1}{2} P^2 + GQ + \braces{\tilde\Omega + \nu I}I - \\
&- \Gamma_K \sqrt{\frac{1}{4}-I^2}\cos\von{KQ-\Delta}, \notag
\end{align}
with
\begin{align}
\tilde \Omega &= \Omega +\frac{A-1}{12}, & \nu &= \frac{2A' - A -1 }{12}\;.
\end{align}
Our understanding of Hamiltonian \eqref{eq:Hpoint} remains essentially intact, therefore, if we introduce a new, $I$-dependent, critical velocity
\begin{align} \label{eq:newPc}
P_\text{c}\von{I} = \frac{\tilde\Omega + \nu I}{K},
\end{align}
which thus changes during the evolution.

The question is whether this straightforward generalization of our Hamiltonian daemon model, or even a more sophisticated generalization that takes into account the deformations of the soliton, can actually approximate the GPE evolution well, when we have general inter-component interactions that break the integrability of the nonlinear Schr\"odinger equation. As an illustration of what can happen in a moderate departure from the simplest case, we again repeat our evolution from the initial state of section II, and with all the same parameters, except that now we use $A=2$ and $A' = 0.5$ instead of $A=A'=1$.
\begin{figure}[thb]
	\centering
	\includegraphics*[trim=105 195 130 225 , width=0.45\textwidth]{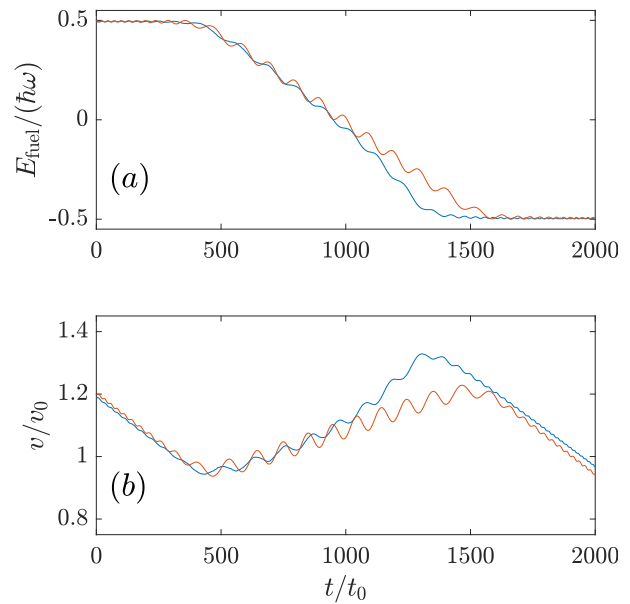}
	\caption{\label{fig:scattering_evolution} Evolution of the fuel energy $E_\text{fuel}/\braces{\hbar \omega}$ (a) and the velocity of the soliton (b), for the case where the critical velocity $P_c(I)$ changes during the evolution. The numerical GPE evolution is plotted in blue, the variational approximation in red. The GPE evolution can be understood quite well from the variational approximation; its error is initially negligible and although it later becomes significant it does not become drastic.}
\end{figure}

In Fig.~\ref{fig:scattering_evolution} we see for this case that the active phase of energy transfer indeed begins when the soliton has decelerated from its high initial velocity to the critical velocity $P_c(I)$ predicted by the variational approximation for the initial fuel level ($I$ close to +1/2). The gradual increase of the active soliton's velocity is also accurately given by the variational approximation, up until about half-way through the plotted evolution. From this point on the variational approximation becomes less accurate as the exactly evolving soliton accelerates more rapidly than the variational model predicts. The variational approximation thus ultimately overestimates the total duration of the active phase, because it underestimates the active soliton's power consumption at its higher final speeds. The variational approximation does correctly predict the final fuel level.

\begin{figure}[thb]
	\centering
	\includegraphics*[trim=110 195 130 225 , width=0.45\textwidth]{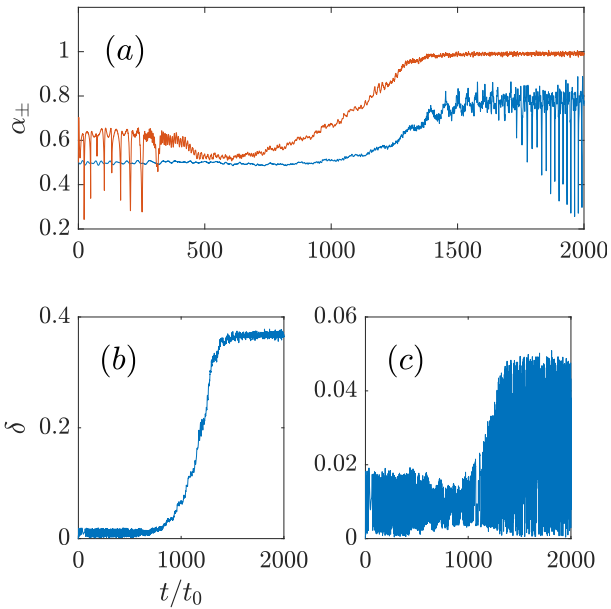}
	\caption{\label{fig:scattering_width} (a) The fit parameters $\alpha_+$ ($\alpha_-$) in blue (red), computed as in Fig.~\ref{fig:alphafromfit} for the case with unequal interaction constants $A=2$ and $A'=0.5$ that was shown in Fig.~\ref{fig:scattering_evolution}. (b) The deviation between the density of the exactly evolving fields, and the soliton form for fixed $\alpha_\pm$, is significantly larger for $t/t_0>1000$ than the error for optimally fitted $\alpha_\pm$, shown in (c).}
\end{figure}

As Fig.~\ref{fig:scattering_width} shows, the onset of discrepancy between the variational model and the GPE evolution is clearly due to increasing distortion of the soliton profile away from the fixed equal widths that the simple variational model assumes. Figure~\ref{fig:scattering_width}(c) indicates that a more general hyperbolic secant form with independently variable $\alpha_\pm$ width parameters should still describe the active soliton well, but the simple equal-width ansatz breaks down. 

For nonintegrable cases like this particular case of $A$ and $A'$ it might therefore be worth pursuing a more general variational approximation with $\alpha_\pm$ as additional variational parameters; we leave this exercise for future work, simply noting that the active bright vector soliton indeed remains qualitatively robust against even significant changes to its nonlinear interaction strengths. Just how far this robustness extends will be our final investigation in this paper.

\subsubsection{(b) Stability range of the active soliton}
For more extreme deviations from $A=A'=1$, solitary wave solutions to the vector nonlinear Schr\"odinger equation may not only deform but even break up, as atoms escape from the solitary wave. Insofar as a solitary wave does still persist, even partially, it is then a related but independent question, whether the solitary wave can move as an active particle by consuming the Zeeman energy of its internal spin depot. Solitary waves in different regimes have long been studied, so our goal in this final portion of our paper is to relate the second question to the first.

To do this we numerically evolved our same initial soliton configuration, until the time $T=5000$ after which any active motion should be complete, under a range of different $A$ and $A'$ cases. In each case we computed how much energy was transferred from spin into motional energy, as well as what fraction of the initial atoms remained in the soliton. We assumed that some kind of active motion must be occurring if any significant fraction of the spin energy is drained; to assess how many atoms remain in the soliton, even if the soliton is distorted, we fit the final mean-field density profile to a hyperbolic secant with independent amplitudes, widths, and centers for the two components. We then defined the fraction of atoms remaining in the bright soliton at $T=5000$---the \emph{retention fraction}---as the Gross-Pitaevskii norm of this fitted hyperbolic secant profile, divided by the exact initial norm. This procedure offers a reasonable measure of how well the `soliton' holds together, as long as the norm which is lost from the soliton tends to spread out in low-density noise that does not significantly affect the best fit $\mathrm{sech}^2$ profile; it would not be accurate if the soliton held together but deformed in shape radically away from $\mathrm{sech}^2$, so that the $\mathrm{sech}^2$ fit represented the surviving solitary wave poorly. Checking particular cases seems to show, however, that this does not occur, and that the $\mathrm{sech}^2$ fitting procedure does provide a good measure of soliton survival.

\begin{figure}[htb]
	\centering
	\includegraphics*[trim=107 245 113 260 , width=0.45\textwidth]{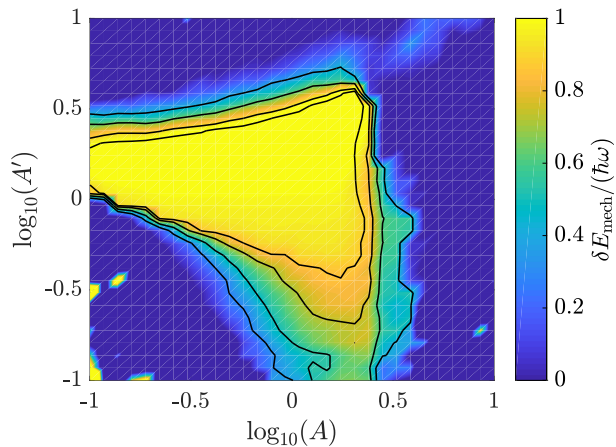}
	\caption{\label{fig:stab} Stability of the soliton and amount of energy transfer for different scattering parameter $A$ and $A'$. The black lines are contours (from outer to inner) of $70\%$, $80\%$, $90\%$, and $95\%$ atom retention by the soliton, as defined in the text. The `soliton' is a stable solitary wave for a wide range of $A$ and $A'$. The color scale linearly indicates the amount of fuel energy that was transferred into mechanical energy over the course of the evolution to $T=5000$. The energy transfer occurs over the whole range in which the soliton is stable.}
\end{figure}

In Fig.~\ref{fig:stab} we show the results in the plane of positive $A$ and $A'$, logarithmically. We indicate soliton survival with black contours of atom retention fraction, and fuel energy consumption with color. The message of the Figure in general is clear: the soliton can essentially remain active as long as it can remain stable at all.

\section{V. Conclusion}

In this paper we have used numerical evolution in Gross-Pitaveskii mean field theory to show that the internal energy of a vector bright soliton in a two-component self-attractive one-dimensional Bose-Einstein condensate can be used to lift the soliton against an external force. The internal energy of the soliton is realized by different chemical potentials for the two components of the Bose-Einstein condensate. The energy transfer from the internal energy to the potential energy is induced by an off-resonant Rabi coupling whose strength varies periodically in space.

This one-dimensional field theory can be approximated at least reasonably well with a variational ansatz of two degrees of freedom. The active-particle behavior of the soliton can be understood from the much simpler Hamiltonian of the variational approximation, which reveals that the spatially periodic Rabi coupling effectively implements the `Hamiltonian daemon' realization of a combustion engine analog, based on a Chirikov resonance. This basic mechanism has been shown to be affected non-trivially by the complications that are possible in the realistic physical representation of the active particle as an atomic bright vector soliton. 

The mechanism has nonetheless been shown to be qualitatively robust in a wide range of conditions. Atomic bright solitons are certainly a highly artificial dynamical system, but they are more complicated than minimal Hamiltonian models: they are complicated enough to be experimentally realizable. Our study can thus be considered as a toy model for the project of extending microscopic understanding of active particles beyond toy models and into real systems.

The authors acknowledge support from State Research Center OPTIMAS and the Deutsche Forschungsgemein- schaft (DFG) through SFB/TR185 (OSCAR), Project No. 277625399.

\end{document}